\documentclass[showpacs, twocolumn]{revtex4}

\usepackage{graphicx}
\usepackage{dcolumn}
\usepackage{amsmath}

\makeatletter
\def\btt#1{\texttt{\@backslashchar#1}}
\DeclareRobustCommand\bblash{\btt{\@backslashchar}} \makeatother

\begin{document}
\title[]{Gravitating magnetic monopole in Vaidya geometry}
\author{Sushant~G.~Ghosh}\email{sghosh.ctp@jmi.ac.in, sgghosh@gmail.com}
\affiliation{Center for Theoretical Physics, Jamia Millia Islamia,
New Delhi - 110025, INDIA}

\author{L. P. Singh}\email{lambodar_uu@yahoo.co.in}
\affiliation{Physics Department, Utkal University, Bhubaneswar 751
004 INDIA}

\begin{abstract}
A magnetic-monopole solution of a non-Abelian gauge theory as
proposed by 't Hooft and Polyakov is studied in the Vaidya
spacetime. We find that the solutions of Einstein equations
generates a geometry of the Bonnor-Vaidya corresponding to
magnetically charged null fluid with Higgs field contributing a
cosmological term.  In the absence of the scalar fields the
corresponding Wu-Yang  solution of the gauge theory still generates
the Bonnor-Vaidya geometry, but with no cosmological term.
\end{abstract}

\pacs{04.20.Jb, 04.70.Bw, 14.80.Hv, 11.15.-q}

\maketitle

\tolerance=5000

\section{Introduction}
One of the most important works on Abelian gauge theories was due to
Dirac, who proposed a solution that corresponds to a pointlike
magnetic monopole with a singularity string running from the
particle's position to infinity \cite{pd}.  This has led to
considerable interest among physicists on the possible existence of
the  magnetic monopole  which was further intensified after 't Hooft
\cite{gtf} has proposed solutions  for a magnetic monopole which
arises as a static solution of the classical equations for
Yang-Mills (YM) field coupled to Higgs fields (see also Polyakov
\cite{amp}). He pointed out that a unified gauge theory in which
electromagnetism is embedded in a semisimple gauge group would
predict the existence of the magnetic monopole as a soliton with
spontaneous symmetry breaking. The  't Hooft-Polyakov magnetic
solutions are similar to the Wu and Yang \cite{wy} classical
monopole solution to the SO(3) isospin gauge theory describing a
magnetic monopole which is pointlike and has a potential which
behaves like $1/r$ everywhere. It behaves like an Abelian theory at
large distance with the gauge-field resembling that due to an
Abelian Dirac magnetic monopole of magnetic charge $1/e$; $e$ is YM
coupling constant.  The question that arises naturally is what
happens when such monopoles are coupled to gravity.  The
generalization of the 't Hooft-Polyakov solutions to curved
space-time also attracted lot of interest
\cite{br,cf,py,nwp,meo,vw}.  Bais and Russell \cite{br} and
independently by Cho and Freund \cite{cf} found a solution to
complete Einstein-Yang-Mills-Higgs (EYMH) system yielding a geometry
of Reissner–Nordstr$\ddot{o}$m-de Sitter with the Higgs field
contributing  to a cosmological term.   Yasskin \cite{py} gave a
explicit algorithm so that from each solution of the
Einstein-Maxwell equations one can set of solutions
 of EYM equations.   A curved-space generalization of the Wu-Yang
 solution is shown to be a special case of  Yasskin's \cite{py} solutions.
 More recently, using Yasskin's \cite{py} procedure, Mazharimousavi and Halilsoy
\cite{shmh07,shmh08,shmh081} have found a sequence of static
spherically symmetric HD-EYM black hole solutions. The remarkable
feature of this Wu-Yang \textit{ansatz} is that the field has no
contribution from gradient and instead has pure YM non-Abelian
component.

It would be interesting to further consider nonstatic generalization
of  't Hooft-Polyakov solutions. It is the purpose of this paper is
to obtain an exact nonstatic solution of the Einstein field
equations for a 't Hooft - Polyakov  solutions in the presence of
the null fluid, i.e.,a nonstatic curved space-time generalization 't
Hooft-Polyakov solutions in Vaidya geometry. The Vaidya geometry
permitting the incorporations of the effects of null fluid offers a
more realistic background than static geometries, where all back
reaction is ignored.  The Vaidya \cite{pc} and Bonnor-Vaidya
\cite{bv} (charged Vaidya) solutions are widely used to model black
hole evaporation and to solve the black hole evaporation problem. It
is also commonly used as a testing ground for various gravitational
scenario and formulation of the cosmic censorship. For a general
method of obtaining spherically symmetric solutions in Vaidya
geometry, see Ref. \cite{ww,dg,gdhd}. The exact solution obtained
represents the generalization, for EYMH system, of the solutions
previously obtained by Bais and Russell \cite{br}, by Cho and Freund
\cite{cf}.  In fact from a physical point of view our solution seems
to be a  more realistic candidates for the study of properties of
collapsing objects.

\section{Basic Equations and Solutions}
We consider  $SO(3)$ gauge theory with structure constant $C_{\left(
\beta\right) \left( \gamma\right) }^{\left( \alpha \right) }$, the
YM fields $F_{a b }^{\left( \alpha\right) }$, triplet of YM field
$A_{a }^{\left( \alpha\right)}$ and a Higgs triplet $\phi^{\left(
\alpha\right)}$ ($a = 0,.\; .\; .\; , 3$ space-time indices and
$\alpha = 1,2,3$ isospace indices).  The standard Einstein-Hilbert
action becomes
\begin{widetext}
\begin{equation}\label{L}
 \mathcal{I} =\int \sqrt{-g} d^4 x \left[ - \frac{1}{2} g^{a b} g^{s d} F^{\left( \alpha\right)}_{a s} F_{b d}^{\left( \alpha\right)
 } - g^{a b}(D_a \phi^{\left( \alpha\right)}) (D_b \phi^{\left(
 \alpha\right)}) -
  \mu^2 (\phi^{(\alpha)}\phi_{(\alpha)}) - \frac{\lambda}{4} (\phi^{(\alpha)}\phi_{(\alpha)})^2 \right] +  \mathcal{I}_N
\end{equation}
\end{widetext}
Here, $g$ = det($g_{ab}$) is the determinant of the metric tensor
and $I_{\mathcal{N}}$ is the action of  null fluid. The classical
equation of motion for YM fields are
\begin{equation}\label{F}
F_{a b }^{\left( \alpha\right) }=\partial _{a }A_{b }^{\left(
\alpha\right) }-\partial _{b }A_{a }^{\left( \alpha\right)
}+\textit{e}\; \epsilon_{\left( \beta\right) \left( \gamma\right)
}^{\left( \alpha \right) }A_{a }^{\left( \beta\right) }A_{b
}^{\left( \gamma \right) }.
\end{equation}%
The gauge covariant derivative $D_a \phi^{\left( \alpha\right)}$ is
\begin{equation}\label{phi}
D_a \phi^{\left( \alpha\right)} =  \partial _{a }\phi^{\left(
\alpha\right)} + \textit{e}\; \epsilon_{\left( \beta\right) \left(
\gamma\right) }^{\left( \alpha \right) }A_{a }^{\left( \beta\right)
}\phi^{\left( \gamma \right) }.
\end{equation}
where $\epsilon_{\left( \beta\right) \left( \gamma\right) }^{\left(
\alpha \right) }$ is totally antisymmetric tensor. Variation of the
action (\ref{L}) with respect to the metric $g_{ab}$, gauge field
$A_{a }^{\left( \alpha\right)}$ and  Higgs field  $\phi^{\left(
\alpha\right)}$ leads to Einstein equation and matter field
equations.  Higgs field vacuum expectation value $|<\phi> | = F$, where
\begin{equation}\label{FS}
    F^2 = -\frac{ 2\mu^2}{\lambda}
\end{equation}
The mass of the Higgs particle is given by $M_H = \sqrt{\lambda} F$.
For the gauge and  Higgs fields we employ the following \textit{ansatz}:
\begin{equation}
 A_{a }^{\left( \alpha \right)} = \epsilon_{a \left(  \beta \right) }^{ b  \left(\alpha \right)} \; \eta_{b} \; r^{\left( \beta
 \right)} A(r), \hspace{.3in} \phi^{\left( \alpha \right)} = r^{\left( \alpha \right)} \phi(r).
\end{equation}
where $\eta_{b} = \delta_b^v $ is timelike unit vector. To construct
spherically symmetric gravitating monopole solution in Vaidya
space-time we employ the Eddington coordinates and adapt the metric
of general spherically symmetric space-time \cite{ww,dg,gdhd}  given
by
\begin{equation}
ds^2 = - A(v,r)^2 f(v,r)\;  dv^2
 +  2 \epsilon A(v,r)\; dv\; dr + r^2  d\Omega^2 \label{metric}
\end{equation}
, where $ d\Omega^2 = d \theta^2 + \sin^2{\theta} d
\phi^2$. Here $A(v,r)$ is an arbitrary function.  It is the field equation
$G^0_1 = 0$ that leads to $ A(v,r) = g(v)$. However, by introducing
another null coordinate $\overline{v} = \int g(v) dv$, we can always
set, without the loss of generality, $A(v,r) = 1$. Therefore the
entire family of solutions we are searching for is determined by a
single function $f(v,r)$.

An exact solution for  $A(r)$ and $\phi(r)$ are determined as

\begin{equation}\label{phis}
   \phi(r) = \frac{F}{r}, \hspace{.3in} \phi^{\left( \alpha \right)} =
 \frac{  r^{\left( \alpha \right)}}{r} F
\end{equation}
and
\begin{equation}\label{as}
A(r) = - \frac{1}{\textit{e}\; r^2}, \hspace{.3in} A_{a }^{\left(
\alpha \right)} = \epsilon_{a \left(  \beta \right) }^{ b
\left(\alpha \right)} \; \eta_{b} \; \frac{r^{\left( \beta
 \right)}}{r} \frac{1}{ \textit{e}\: r}.
\end{equation}
That such a solution corresponds to a magnetic monopole can be seen by inserting it
into gauge-invariant generalization of the electromagnetic field
tensor:
\begin{equation}\label{eft}
\mathbf{F}_{ab} =\frac{ \phi_{\left( \alpha \right)}}{|\phi|} F_{a b
}^{\left( \alpha\right) } - \frac{1}{e} \;
\epsilon_{(\alpha)(\beta)( \gamma)} \frac{ \phi^{\left( \alpha
\right)}}{|\phi|} (D_a \phi^{\left( \beta\right)}) (D_b \phi^{\left(
 \gamma \right)})
\end{equation}
which yields
\begin{equation}
\mathbf{F}_{ab} = - \epsilon_{a b (\alpha)}\frac{ r^{(\alpha)}}{r^3}
\end{equation}
Clearly $\mathbf{F}_{ab}$ satisfies Maxwell equations, except at
$r=0$, and corresponds to magnetic field
\begin{equation}
    \overrightarrow{B} = Q \frac{\overrightarrow{r}}{r^3}
\end{equation}
of a magnetic point charge with $Q=1/\textit{e}$.  In this letter, without loss of generality, we choose $Q=Q(v)$.
It is seen that, for the metric  (\ref{metric}), the matter field equation admits solution $Q(v)=1/\textit{e}$.
The Einstein equations is%
\begin{equation}\label{eq:ee}
G_{a b }=T_{a b}.
\end{equation}%
Expressing the total energy-momentum  tensor (EMT)  as
\begin{equation}\label{emt}
T_{a b} =T_{a b }^G+T_{a b }^N,
\end{equation}
where  the gauge EMT  $T_{a b }^G$ is
\begin{widetext}
\begin{eqnarray}
T_{a b }^G & = & 2\Big[ g^{mn} F^{\left( \alpha\right)}_{a m} F_{b
n}^{\left( \alpha\right)} -\frac{1}{4} g_{ab}g^{mn}g^{s t} F^{\left(
\alpha\right)}_{m s} F_{n t}^{\left( \alpha\right)} + (D_a
\phi^{\left( \alpha\right)}) (D_b \phi^{\left(
 \alpha\right)})  \\ & & -  \frac{1}{2}g_{a b}g^{m n}(D_m \phi^{\left( \alpha\right)}) (D_n \phi^{\left(
 \alpha\right)}) - g_{a b} \left( \frac{1}{2}\mu^2 (\phi^{(\alpha)}\phi_{(\alpha)}) + \frac{\lambda}{8} (\phi^{(\alpha)}\phi_{(\alpha)})^2
 \right)
 \Big]
\end{eqnarray}
\end{widetext}
and the null fluid EMT is
\begin{equation}\label{emtn}
T_{a b}^N = \psi(v,r) l_a l_b
\end{equation}
with $\psi(v,r)$, the nonzero energy density and $l_a$ is a null
vector such that $l_{a} = \delta_a^0, l_{a}l^{a} = 0.$ For $a \neq
b$, $T^a_b=0$ except for a nonzero off diagonal component $T^r_v$.
It may be recalled that EMT of a Type II fluid has a double null
eigenvector, whereas an EMT of a Type I fluid has only one timelike
eigenvector \cite{he}. In addition, we observe that the metric
(\ref{metric}) requires that $T_v^v = T_r^r$.

Inserting Eqs.~(\ref{phis}) and(\ref{as}), we obtain the expression
for the gauge-field tensor which in spherically coordinates becomes

\begin{eqnarray}
F_{\theta \phi }^{(x)} &=& -Q(v) \sin^2 \theta\; \cos \phi   \\
F_{\theta \phi }^{(y)} &=&   Q(v) \sin^2 \theta \; \sin  \phi  \\
F_{\theta \phi }^{(z)} &=&  -Q(v) \sin \theta \; \sin  \phi
\end{eqnarray}
all other components vanish.  Putting this in expression for EMT and
setting $\beta = \mu^4/ \lambda$, the EMT can be written
as:
\[
T^a_b = \left(%
\begin{array}{cccc}
-\frac{Q^2(v)}{r^4} +\beta \; & 0 \;  & 0 \;  & 0 \; \\

\psi(v,r) & - \frac{Q^2(v)}{r^4} +\beta   & 0 & 0 \\
   0 & 0 &\frac{Q^2(v)}{r^4} +\beta& 0 \\
   0 & 0 & 0 & \frac{Q^2(v)}{r^4} +\beta \\
\end{array}%
\right).
\]
Here $\beta$ is contribution from scalar field due to spontaneous
symmetry breaking.  In the limit,  $\beta =0$, EMT is same as in
charged null fluid because $D_a \phi^{\alpha}$ vanish everywhere and
does not contribute to EMT.

For the EMT (\ref{emt}) and with the metric (\ref{metric}), the
Einstein equations (\ref{eq:ee}) reduce to:

\begin{subequations}
\label{fe1}
\begin{eqnarray}
\psi= - \frac{1}{r} \frac{\partial f(v,r)}{\partial v},
\label{equationa}\\
\frac{1}{r}\frac{\partial f(v,r)}{\partial r} - \frac{1}{r^2} +
\frac{f(v,r)}{r^2} = -\frac{Q^2(v)}{r^4} +\beta,
\label{equationb} \\
\frac{1}{2}\frac{\partial^2 f(v,r)}{\partial r^2} +
\frac{1}{r}\frac{\partial f(v,r)}{\partial r} = \frac{Q^2(v)}{r^4}
+\beta  \label{equationc}
\end{eqnarray}
\end{subequations}

It may be noted that in view of Eq.~(3), the gauge field has only
the angular components nonzero and they go as $r^{-2}$ which in turn
makes $T_{ab}^G$ go as $r^{-4}$. The null fluid part will be given
by $T^r_v = \psi(r,v)$. The last two equations are not independent
and it suffices to integrate Eq.~(\ref{equationb}) to give
\begin{equation}
f(v,r) =  1 - \frac{2 M(v)}{r} + \frac{Q^2(v)}{r^2} + \beta
\frac{r^2}{3} \label{eq:sol}
\end{equation}
where $M(v)$ is an arbitrary function of $v$.  Since YM $T_{ab}^G$
go as $r^{-4}$ (the same as for Maxwell field in $N=4$),
 that is why its contribution in $f$ as in 4-dimensional
Reissner-Nordstr$\ddot{o}$m static or Bonnor-Vaidya radiating
black-hole \cite{gd}. Thus the metric describing the in $(v,r,
\theta, \phi)$ coordinates reads as:

\begin{eqnarray}
ds^2 = - \left(1 - \frac{2 M(v)}{r} + \frac{Q^2(v)}{r^2} + \beta
\frac{r^2}{3}\right) \;  dv^2 \nonumber \\
 +  2 \; dv\; dr + r^2  d\Omega^2 \label{sol1}
\end{eqnarray}
From Eq.~(\ref{equationa}), we obtain the energy density of the null
dust with gauge charge as
\begin{eqnarray}
\psi(v,r) =  \frac{2}{r^2} \frac{dM(v)}{dv} - \frac{2Q(v)}{r^3}
\frac{dQ(v)}{dv} \label{density}
\end{eqnarray}
and YM energy density and transverse stress are given by
\begin{eqnarray}
  \zeta(v,r) &=& \frac{Q^2(v)}{r^4} - \beta  \\
  P(v,r) &=& \frac{Q^2(v)}{r^4} + \beta  \label{denpr}
\end{eqnarray}
 The family of solutions discussed
here, in general, belongs to Type II fluid defined in \cite{he}.
These are same results as one would expect for charge null dust in
the Abelian theory, i.e., the geometry is precisely of the
Bonnor-Vaidya-de Sitter \cite{bv,lz} form and the charge that
determines the geometry is YM gauge charge, whereas the Higgs field
$\beta$ playing the role of a cosmological constant, from a formal
mathematical point of view.  Thus we can also say that any solution
of the Einstein-Maxwell
 system is also solution of the Einstein-Yang-Mills (EYM) system.

The Kretschmann scalar ($K = R_{abcd} R^{abcd}$, $R_{abcd}$ is the
Riemann tensor) for the metric (\ref{sol1}) reduces to
\begin{equation}
K = \frac{48}{r^6} \left[M^2(v) - \frac{2}{r} Q^2(v)M(v) +
\frac{7}{6}
 \frac{Q^4(v)}{r^2}  \right] + \frac{8}{3} \beta^2     \label{eq:ks}
\end{equation}
So the Kretschmann scalar diverges along $r = 0$. The Weyl scalar
($C = C_{abcd} C^{abcd}$, $C_{abcd}$ is the Weyl tensor) reads
\begin{eqnarray}
C  = \frac{48}{r^6} \left[M^2(v) - \frac{2 M(v) Q^2(v)}{r} +
\frac{Q^4(v)}{r^2} \right]            \label{eq:ws1}
\end{eqnarray}
which also diverges along $r=0$.

In the rest frame associated with the
observer, the energy-density of the matter will be given by,
\begin{equation}
\psi = T^r_v,\hspace{.1in} \zeta = - T^t_t = - T^r_r
=\frac{Q^2(v)}{r^4} - \beta   , \label{energy}
\end{equation}
 and the principal pressures are $P_i =
T^i_i$ (no sum convention). \\

 \noindent \emph{a) The weak energy
conditions} (WEC): The EMT obeys inequality $T_{ab}w^a w^b \geq 0$
for any timelike vector, i.e.,
\begin{equation}
\psi \geq 0,\hspace{0.1 in}\zeta \geq 0,\hspace{0.1 in} P_{\theta}
\geq 0, \hspace{0.1 in} P_{\phi} \geq 0. \label{wec}
\end{equation}
We say that strong energy condition (SEC), holds for Type II fluid
if, Eq.~(\ref{wec}) is true., i.e., both WEC and SEC, for a Type
II fluid, are identical. \\

\noindent {\emph{b) The dominant energy conditions }}: For any
timelike vector $w_a$, $T^{ab}w_a w_b \geq 0$, and $T^{ab}w_a$ is
nonspacelike vector, i.e.,
\begin{equation}
\psi \geq 0,\hspace{0.1 in}\zeta \geq P_{\theta}, P_{\phi} \geq 0.
\hspace{0.1 in}
\end{equation}

Clearly, $(a)$ is satisfied if $Q^2(v) \geq \beta r^4. $  However,
$\psi
> 0$ gives the restriction on the choice of the functions $M(v)$
and $Q(v)$. From Eq.~(\ref{energy}),  we observe $\psi
> 0$ requires,
\begin{equation}
\frac{2}{r^2} \frac{dM(v)}{dv} > \frac{2Q(v)}{r^3} \frac{dQ(v)}{dv}
\end{equation}
 We note that the stress tensor in general may not obey the
weak energy condition.  In particular, if $dM/dQ > 0$ then there
always exists a critical radius $r_{c} = Q \dot{Q}/ \dot{M}$ such
that when $r < r_{c}$ the weak energy condition is always violated.
On the other hand, the DEC may not hold.

Thus, we have constructed an explicit nonstatic magnetically charged
null fluid solutions of a non-Abelian gauge theory coupled to
gravitation and Higgs field. Thus we have exact Vaidya like
solutions of the Einstein-Yang-Mills-Higgs model.  This yields same
results as one would expect for a charge null fluid in the Abelian
theory, i.e., the geometry is precisely of the Bonnor-Vaidya form
and the charge that determines the geometry is magnetic charge
($1/e$) with Higgs field contributing to  cosmological constant like
term.  The geometry becomes asymptotically flat in the absence of
the scalar field.  Thus we found generalization of the model
discussed previously by Bais and Russell \cite{br} and independently
by Cho and Freund \cite{cf}. Now a few comments are in order: (1) If
one adds cosmological term to the Lagrangian, then the solution
(\ref{eq:sol}) requires additional term:
\begin{equation}
f(v,r) =  1 - \frac{2 M(v)}{r} + \frac{Q^2(v)}{r^2} - (\Lambda -
\beta)\frac{r^2}{3} \label{eq:sol2}
\end{equation}
It may be noted that  Eq.~(\ref{eq:sol2}) has three types of
solutions, namely, Bonnor-Vaidya-de Sitter, Bonnor-Vaidya  and
Bonnor-Vaidya-anti-de Sitter solutions depending on whether $\Lambda
> \beta$, $\Lambda = \beta$, or $\Lambda < \beta$ respectively. (2)
Further, when the magnetic charge is switched off, i.e. if $ Q^2(v)
= 0 $, then, from Eq.~(\ref{sol1}), one obtains Vaidya-de Sitter
metric and if both $Q^2(v) = \beta = 0 $ one gets Vaidya metric. (3)
Another case is obtained if both $M=Q=$ constant. By introducing the
transformation
\begin{equation}
dt = dv - \left(1 - \frac{2 M}{r} + \frac{Q^2}{r^2} + \beta
\frac{r^2}{3} \right)^{-1} dr
\end{equation}
Eq.~(\ref{sol1}) becomes
\begin{eqnarray}
ds^2 = - \left(1 - \frac{2 M}{r} + \frac{Q^2}{r^2} + \beta
\frac{r^2}{3} \right) dt^2 \nonumber \\ + \left(1 - \frac{2 M}{r} +
\frac{Q^2}{r^2} + \beta \frac{r^2}{3} \right) ^{-1}
 dr^2   + r^2  d\Omega^2 \label{sol2}
\end{eqnarray}
which is the same as that given by Bais and Russell \cite{br} and
Cho and Freund \cite{cf}. Thus, as mentioned earlier, several known
models can be recovered from our analysis.
\section{Concluding remarks}
Because of the complex nature of the full Einstein equations, the
metrics with special symmetries are used to construct gravitational
collapse models. One such case is the two-dimensional reduction of
general relativity obtained by imposing spherical symmetry. Even
with this reduction, however, very few inhomogeneous exact nonstatic
solutions have been found. One well-known example is the Vaidya
metric\cite{pc}.  In view of this, the solutions presented here can
be useful to get insights into more general gravitational collapse
situations and to model the dynamical evolution of a Hawking
evaporating black holes. It would be useful to investigate the back
reaction of an evaporating black hole with magnetic-monopoles and
also the changes in the structure and location of the horizons when
a black hole with magnetic monopole radiates.  Our solution can be
also utilized to study Vaidya collapse with a magnetic monopole
field and examine the formation of black holes and naked
singularities and examine how the perturbation induced by the
external matter fields affect the formation or otherwise of the
naked singularity. The relevant question is whether the effect of
such external fields could remove the occurrence of the same.
Indeed,  one can claim that the
 space-time discussed here has same singularity behavior as
the Bonnor-Vaidya-de Sitter \cite{lz} due to mathematical similarity
of our solution with that of Bonnor-Vaidya-de Sitter. For
Bonnor-Vaidya-de Sitter case,  shell focusing strong curvature naked
singularities do arise and hence in our model as well the naked
singularity definitely develops. In that sense, the gauge charge
does not remove the naked singularity, which thus displays stability
with respect to this particular mode of perturbation.  The
usefulness of these models is that  they do offer opportunity to
explore of properties of singular space-time and, in the case of
curvature singularity to address issue such as local or global
nakedness and strength. These and other related topics are under
investigation \cite{gd}.

It also serves to illustrate the much richer interplay that can
occur among particle physics and general relativity when more
involved theoretical models are considered. As a final remark, it
would be also interesting to see how the results get modified in
higher dimensional space-time with the Gauss-Bonnet combination of
quadratic curvature terms \cite{gd}.

\section*{Acknowledgments}
One of the authors SGG is supported by  university grant commission
(UGC) major research project grant F. NO. 39-459/2010 (SR).

\end{document}